\pdfoutput=1
\documentclass[10pt,conference]{IEEEtran}
\IEEEoverridecommandlockouts

\usepackage{cite}
\usepackage{url}
\usepackage{graphicx}
\usepackage{microtype}
\usepackage{xspace}
\usepackage{balance}
\usepackage{subcaption}

\usepackage{amsmath,amssymb,amsfonts,amsthm}
\usepackage[mathscr]{eucal}
\usepackage{mathrsfs}

\usepackage{booktabs}
\usepackage{multirow}
\usepackage{makecell}
\usepackage{array}
\usepackage{tabularx}
\usepackage{threeparttable}
\usepackage[table]{xcolor}

\usepackage{float}
\usepackage{wrapfig}

\usepackage[most]{tcolorbox}
\newcommand{\rqbox}[1]{
\begin{tcolorbox}[tile, size=fbox, boxsep=2mm, boxrule=0pt, top=0pt, bottom=0pt,
borderline west={1mm}{0pt}{blue!50!white}, colback=blue!5!white]
#1
\end{tcolorbox}
}
\usepackage{enumitem}
\usepackage{setspace}

\usepackage{pifont}
\usepackage{bbding}
\usepackage{utfsym}

\definecolor{cmarkfg}{RGB}{34,139,34}
\definecolor{xmarkfg}{RGB}{180,30,30}

\newcommand{\CM}{%
  \textcolor{cmarkfg}{\ding{51}}}

\newcommand{\XM}{%
  \textcolor{xmarkfg}{\ding{55}}}

\usepackage{tikz}
\usetikzlibrary{
  shapes.geometric,
  arrows.meta,
  positioning,
  calc,
  fit,
  backgrounds,
  decorations.pathreplacing
}

\usepackage[colorlinks,bookmarksopen,bookmarksnumbered,citecolor=green]{hyperref}

\usepackage{booktabs}
\usepackage{multirow}
\usepackage{colortbl}
\usepackage{xcolor}
\usepackage{pifont}
\usepackage{array}
\usepackage{placeins}   % 提供 \FloatBarrier

\usepackage{adjustbox}
\usepackage[table]{xcolor}
\usepackage{booktabs}
\usepackage{multirow}
\usepackage{array}

\definecolor{cmarkfg}{RGB}{34,139,34}
\definecolor{xmarkfg}{RGB}{180,30,30}
\definecolor{rowshade}{RGB}{245,247,252}
\definecolor{totalbg} {RGB}{232,240,254}
\definecolor{syshead} {RGB}{210,228,248}

\newif\ifshowcomments
\showcommentstrue

\newcommand{\sysname}{\textsc{KeaRepair}\xspace}

\newcommand{\code}[1]{{\fontfamily{cmtt}\fontseries{m}\fontshape{n}\selectfont\small{#1}}}

\def\BibTeX{{\rm B\kern-.05em{\sc i\kern-.025em b}\kern-.08em
    T\kern-.1667em\lower.7ex\hbox{E}\kern-.125emX}}

\begin{document}

\title{Knowledge-Enhanced Agentic Vulnerability Repair}

\author{
    \IEEEauthorblockN{Sicong Cao$^\dagger$, Hao Ma$^\dagger$, Le Yu$^\dag$\textsuperscript{\Envelope}, Kangyi Ding$^\ddagger$, Xiaolei Liu$^\ddagger$, Terry Yue Zhuo$^\S$, Bo Wang$^\P$, \\ 
    Xingwei Lin$^\diamondsuit$, Xiaobing Sun$^\spadesuit$, Linzhang Wang$^\clubsuit$, David Lo$^\heartsuit$}
    \IEEEauthorblockA{$^\dagger$Nanjing University of Posts and Telecommunications\quad $^\ddagger$National Interdisciplinary Research Center of Engineering Physics\\ $^\S$Alibaba Qwen\quad $^\P$Beijing Jiaotong University\quad $^\diamondsuit$Zhejiang University\quad $^\spadesuit$Yangzhou University\quad $^\clubsuit$Nanjing University\\ $^\heartsuit$Singapore Management University}
    % \IEEEauthorblockA{$^\dag$\{DX120210088, xbsun, xiaoxuewu, lilibo, lb\}@yzu.edu.cn,}
    % \IEEEauthorblockA{$^\ddag$\{hb187361, yu.oyy, chuanlei.mchl, jiajia.lijj, lenx.wei\}@antgroup.com,}
    % \IEEEauthorblockA{$^\S$chaoz@tsinghua.edu.cn, $^\P$tsu@sei.ecnu.edu.cn}
}

\maketitle

\begin{abstract}
Frontier foundation models have significantly accelerated vulnerability discovery while slashing costs, but the bigger challenge is how the remediation side keeps up. Despite recent progresses in Automated Vulnerability Repair (AVR), current solutions struggle to reliably pinpoint vulnerability causes, and insufficiently utilize the prior fix knowledge to guide the patch generation process, undermining their effectiveness in practice.

To address this gap, we propose \sysname, a novel agentic AVR approach that grounds patch generation in verified program facts and high-level vulnerability knowledge. Specifically, \sysname first extracts dual-view multi-dimensional knowledge items from historical vulnerability-patch pairs, and constructs dedicated retrieval knowledge bases. It then employs a tool-augmented agent that performs ReAct-style reasoning to collect verified program facts for vulnerability diagnosis. Finally, based on the diagnostic results, \sysname performs knowledge-level retrieval-augmented patch generation and iteratively refines patches through a closed-loop validation process. Experimental results show that \sysname significantly outperforms existing AVR approaches on 55 reproducible C/C++ vulnerabilities. When paired with Gemini-3.1-Pro, KeaRepair successfully repairs 46 vulnerabilities, achieving a repair rate of 83.64\%. Moreover, \sysname fixes nine unique vulnerabilities that the state-of-the-art baseline PatchAgent cannot address, and further demonstrates strong cross-language generalizability.

\end{abstract}

\section{Introduction}
As the cybersecurity capabilities of frontier foundation models like Claude Mythos \cite{mythos} and GPT-5.5-Cyber \cite{5.5-cyber} continue to evolve, the cost and effort required to find and exploit software vulnerabilities have all dropped dramatically. As of May 2026, the public Common Vulnerabilities and Exposures (CVE) program \cite{CVE} has disclosed 29,120 new vulnerabilities, an increase of approximately 39.89\% year-over-year. The resulting backlog far outpaces the resources available to triage them, creating a strong demand for tools that can quickly generate correct patches for known vulnerabilities \cite{DBLP:conf/uss/0095SW0025}.

\noindent\textbf{Existing Efforts.}
Recent progresses in Large Language Models (LLMs) have sparked interest in their application to Automated Vulnerability Repair (AVR) \cite{DBLP:journals/tosem/ZhouCSL25}, primarily due to their superior reasoning capabilities compared to traditional rule-intensive \cite{senx,vulnfix} or popular data-driven approaches \cite{vulrepair,vulmaster}. Early attempts primarily follow the one-step repair paradigm, employing prompt engineering \cite{DBLP:conf/sp/PearceTAKD23}, reinforcement learning \cite{DBLP:conf/kbse/WenLYGY25,DBLP:journals/corr/abs-2510-01002}, or Retrieval-Augmented Generation (RAG) \cite{appatch} to directly output candidate patches. Afterwards, a series of LLM-based patching agents \cite{looprepair,patchagent,san2patch} have been proposed to autonomously localize faults and synthesize patches in an iterative workflow.

\noindent\textbf{Limitations.}
While promising, existing LLM-based approaches suffer from two fundamental \textbf{L}imitations.

\underline{\emph{L1: Unreliable Root Cause Analysis.}}
Identifying the root cause of a vulnerability and providing an optimal code location to apply patches is typically the gold standard in most AVR approaches \cite{vulrepair,vulmaster,appatch}. However, this assumption does not align with the reality faced by developers: even the state-of-the-art localization tool struggles to precisely pinpoint the exact statements that need to be fixed \cite{DBLP:conf/uss/Hu0SGZXY025}. More importantly, root causes often do not intersect with the fix location, in fact they may be far apart \cite{DBLP:conf/icse/SoremekunKBP23}. Instead, agentic AVR approaches \cite{patchagent,san2patch} delegate localization entirely to the LLM. Given a sanitizer report, the patching agent reasons about the vulnerability, supplemented by the context retrieved from the project codebase. Unfortunately, due to the inherent hallucination issue \cite{DBLP:journals/corr/abs-2410-09997}, solely relying on the internal knowledge and free-form reasoning of LLMs is not enough. Additionally, the demand for sanitizer report that only applicable to memory safety issues significantly constrains their generalizability.

\underline{\emph{L2: Insufficient Utilization of Vulnerability knowledge.}}
Despite powerful reasoning capabilities, LLMs have been proven to still require some guidance in orchestrating the steps for fixing \cite{yin2024thinkrepair,DBLP:conf/kbse/GaoWGWZL23}. A straightforward countermeasure is RAG, which dynamically retrieves a set of similar vulnerability-patch pairs as references, as prior works \cite{looprepair,FixRepair} do. Nevertheless, naive retrieval strategies are suboptimal as they primarily focus on surface-level code similarity, neglecting the deeper cause-level logical similarities that are critical in patch generation \cite{Vulkey}. Although APPatch \cite{appatch} alleviates this problem to some extent by prompting the LLMs to select historical exemplars that closely match the root cause of the sample to be patched, it still suffers from the precision and scalability issues.

% The first limitation concerns the retrieval mechanism. The root cause of a real-world vulnerability distributes repair-relevant signals unevenly across multiple semantic dimensions---data flow, control flow, and API semantics. Existing RAG-based methods construct a single holistic root-cause description as the retrieval query, which is inevitably dominated by the strongest surface signal while leaving repair-critical signals along other dimensions underrepresented. As a result, the retrieved exemplar matches the dominant surface pattern but differs fundamentally in fix location and repair strategy, biasing the LLM toward an incorrect repair paradigm. For example, on CVE-2017-5969 in \texttt{libxml2}, a representative RAG-based system retrieves a structurally dissimilar CWE-476 exemplar driven by the null-dereference surface signal, and generates a null guard at the wrong location---hundreds of lines away from the actual bug site---while the correct fix requires a missing bound check at function entry, a signal residing entirely in the control-flow and API-semantic dimensions. 

\noindent\textbf{Our Work.}
To tackle the above two challenges, we propose \sysname, a novel \textbf{K}nowledge-\textbf{E}nhanced \textbf{A}gentic vulnerability \textbf{R}epair approach. The key insights underlying \sysname include: (\ding{182}) reliable vulnerability causes should be derived from verified program facts rather than purely static LLM reasoning, as well as (\ding{183}) high-level vulnerability knowledge from historical fixes helps filter out superficially similar exemplars with divergent repair patterns. Specifically, \sysname consists of three tightly coupled phases. \textbf{First}, inspired by that most vulnerabilities can be modeled from the view of control- or data-flow, \sysname characterizes and distills view-specific multi-dimensional vulnerability knowledge under the supervision of ground-truth patches, with dedicated knowledge bases for exemplar retrieval. \textbf{Second}, instead of assuming perfect localization or asking an LLM to infer the vulnerability cause from incomplete symptoms, \sysname leverages a set of specialized static analysis tools to actively gather verified program facts from the vulnerable code, thereby minimizing inherent hallucinations of LLMs. \textbf{Third}, with the structured diagnostic report, \sysname constructs view-specific semantic queries to retrieve relevant historical exemplars as well as their corresponding knowledge items, and synthesizes candidate patches.

\noindent\textbf{Evaluation.}
We implement a prototype system of \sysname, and conduct comparative experiments with eight state-of-the-art AVR approaches \cite{vrepair,vulmaster,vulrepair,vqm,vulnfix,extractfix,appatch,looprepair} on 55 reproducible C/C++ vulnerabilities sourced from \textsc{Vul4C} \cite{DBLP:conf/uss/Hu0SGZXY025}. Experimental results show that \sysname substantially outperforms all studied approaches. In particular, when paired with Gemini-3.1-Pro, \sysname achieves the optimal cost-benefit trade-off, repairing 83.64\% (46/55) of vulnerabilities at a lower cost (\$0.30) and within practical timeframes (under 50 seconds) per vulnerability. More importantly, \sysname is able to fix nine unique CVEs that the best-performing baseline PatchAgent cannot address. In addition, \sysname demonstrates remarkable generalizability to more complex vulnerabilities in other programming languages \cite{DBLP:journals/corr/abs-2511-11019}, yielding a success rate of up to 85.55\% (148/173).

\noindent\textbf{Contributions.}
This paper makes the following contributions:

\begin{itemize}[leftmargin=1em]
  \item \textbf{New Dimension.}
  We identify and address two key challenges of existing LLM-based AVR approaches: unreliable root cause analysis and insufficient knowledge utilization.
  
  \item \textbf{Novel Approach.}
  We propose \sysname, a novel agentic AVR approach. By grounding patch generation in verified program facts and multi-dimensional vulnerability knowledge, \sysname effectively alleviates the inherent hallucination issue of purely static LLM reasoning while sufficiently unleashing their in-context learning capabilities. 

  \item \textbf{Extensive Study.}
  We evaluate \sysname against a rigorous benchmark of 55 reproducible C/C++ vulnerabilities. The experimental results demonstrate that \sysname achieves superior performance across all vulnerabilities types by leveraging diverse LLM backbones, significantly outperforming prior approaches. The extensive cross-language evaluation further reveals the strong generalizability of \sysname.

\end{itemize}

% \noindent\textbf{Data Availability.}
% All scripts and data to reproduce our experiments are publicly available at \url{https://anonymous.4open.science/r/KEARepair-D155}.

\section{Background and Related Work}\label{related work}

\noindent\textbf{Automated Vulnerability Repair.}
Research on Automated Vulnerability Repair (AVR) has witnessed significant progress over the past two decades \cite{DBLP:conf/uss/0095SW0025}, evolving from search-based and template-guided to more intelligent approaches. One such approach is Senx \cite{senx}, which uses heuristic rules to identify safety properties violated by vulnerabilities, and then generates predicates for patch synthesis. Another notable approach is ExtractFix \cite{extractfix}, which employs symbolic execution to derive crash-free constraints from a PoC and transforms them to candidate patches, without relying on extensive test suites.

As vulnerabilities grow in volume and complexity, a series of learning-based approaches~\cite{vrepair,vulrepair,vqm,vulmaster} have been proposed to improve automation by learning repair patterns from historical vulnerability-patch pairs. They typically frame the task as a sequence-to-sequence problem and operate at the function-level. For example, VulRepair~\cite{vulrepair} incorporates pre-trained CodeT5 to learn better code representations for patch generation, while VulMaster~\cite{vulmaster} augments its fine-tuning dataset with CWE knowledge to further improve performance. Most recently, LLM-based approaches have shown promise due to the ability to generalize from a vast amount of data~\cite{DBLP:journals/tosem/ZhouCSL25}. Pearce et al. \cite{DBLP:conf/sp/PearceTAKD23} systematically examines LLMs' patch generation capability in zero-shot settings, highlighting that they still struggle to generate plausible fixes in real-world scenarios. San2Patch~\cite{san2patch} employs multi-stage reasoning with LLMs to generate patches using only sanitizer logs and source code.

\noindent\textbf{Software Engineering Agents.}
LLM agents, hereinafter also referred to as \emph{agents} for short, are AI systems that autonomously execute complex tasks by equipping LLMs with the capabilities of perceiving and utilizing external resources and tools. In recent years, a number of agent-driven solutions have been proposed to complete real-world software engineering tasks \cite{survey}, such as GitHub issue resolution \cite{DBLP:conf/issta/0002RFR24}. For example, SWE-agent \cite{DBLP:conf/nips/YangJWLYNP24} designs an Agent-Computer Interface (ACI) to bridge LLMs and terminal environments via structural tool interactions, while Atomizer~\cite{Atomizer} introduces a collaborative multi-agent framework to untangle composite commits by inferring the intent of each code change and iteratively self-correcting.

\begin{figure*}[htbp]
  \centering
  \includegraphics[width=.85\linewidth]{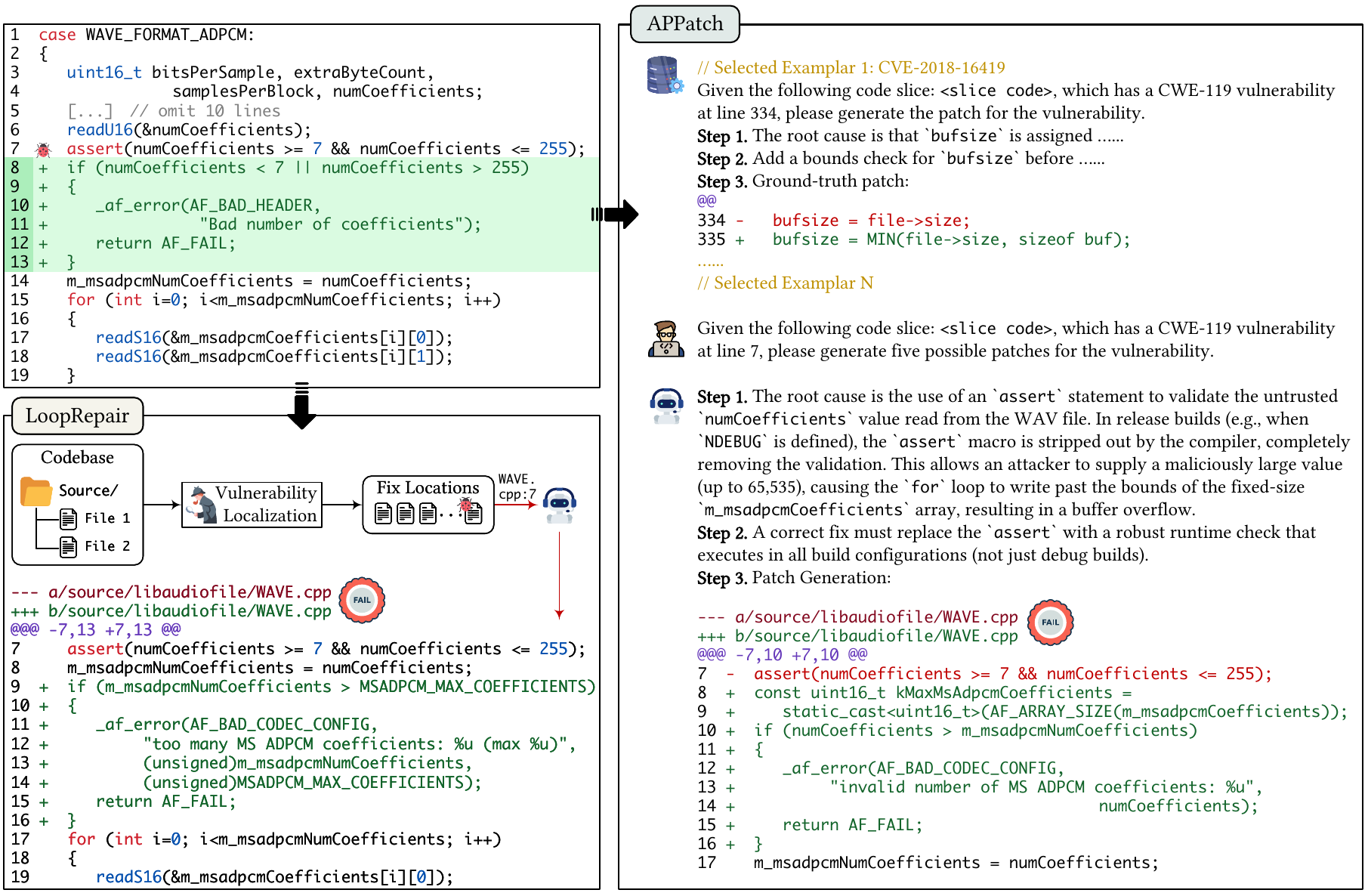}
\caption{An illustrative example using a real-world vulnerability (CVE-2017-6828) from the \code{audiofile} project. Note the patch is simplified to include only the relevant context for brevity.}
\label{fig:motivation}
\vspace{-4mm}
\end{figure*}

\section{Motivation}\label{Motivation}
To illustrate the fundamental limitations of existing LLM-based AVR approaches and motivate the key insights of our approach, we use a real-world vulnerability CVE-2017-6828 \cite{CVE-2017-6828} from \code{audiofile} as a running example.

\subsection{Motivating Example}
As shown in Fig.~\ref{fig:motivation} (left), in the case of \code{WAVE\_FORMAT\_ADPCM}, a 16-bit unsigned integer \code{numCoefficients} is read directly from an untrusted WAV file (line 280) to determine how many iterations should perform. To avoid the loop from executing far more times than the array \code{m\_msadpcmCoefficients} can accommodate, an assert statement is applied (line 7) to validate whether the value of \code{numCoefficients} falls within an expected range (between 7 and 255). Nevertheless, asserts are typically eliminated when building the library with \code{NDEBUG}. As a result, the intended input validation is completely absent in production environments, introducing a heap-based buffer overflow vulnerability that allows an attacker to supply any value up to 65,535 that past the bounds of \code{m\_msadpcmCoefficients}. To fix this vulnerability, developers add a robust runtime check that executes in all build configurations (lines 8-13) to ensure the number of \code{coefficients} is within the array range.

\rqbox{\ding{43} \textbf{Observation~1}
$\blacktriangleright$
Symbolic localization tools or agentic reasoning alone struggles to reliably identify root causes.
$\blacktriangleleft$}

As a prerequisite for initiating the patching process, vulnerability localization aims to find a program point at which the \emph{root cause} of an observed program failure can be fixed \cite{DBLP:conf/asiaccs/ShenKDSR21}. Similar to prior AVR approaches \cite{vulnfix,vulrepair}, LLM-based solutions typically assume perfect vulnerability localization, where the vulnerable function and all locations for modification are given \cite{Vulkey,appatch}. However, this assumption is unrealistic in practice as developer-provided fix locations are unavailable \cite{DBLP:conf/icsm/KabadiKXBPLL023}. To tackle this issue, LoopRepair \cite{looprepair} employs CrashAnalysis \cite{DBLP:journals/tosem/ShariffdeenTNGR25}, a state-of-the-art localization tool based on concolic execution, to identify root cause locations, and instructs the LLM to predict the corresponding fix locations. While effective, such a dynamic solution may suffer from significant scalability and precision issues. As shown in Fig. \ref{fig:motivation} (bottom), except the actual root cause statement (line 7), LoopRepair mistakenly pinpoints 13 faulty locations across six files, with a false positive rate of 92.86\%. Another common paradigm is agentic reasoning, which prompts the LLM to autonomously analyze the crash log, localize the issue, and generate patches without human intervention \cite{patchagent,san2patch}. However, they are limited to generating patches for vulnerabilities detected by a sanitizer.

\rqbox{\ding{43} \textbf{Observation~2}
$\blacktriangleright$
Prior vulnerability knowledge should be better utilized to guide the patch generation process.
$\blacktriangleleft$}

Apart from unreliable root cause analysis, Another limitation of existing LLM-based AVR approaches is their insufficient utilization of vulnerability knowledge. As reported by Pearce et al. \cite{DBLP:conf/sp/PearceTAKD23}, LLMs struggle to generate plausible fixes in a zero-shot setting. An intuitive way is RAG, which feeds LLMs with a few relevant demonstrations to provide valuable guidance \cite{looprepair,DBLP:conf/kbse/GaoWGWZL23}. For example, PailGen \cite{FixRepair} dynamically retrieves top-ranked vulnerability-patch pairs that are both lexically and semantically similar to the testing vulnerable code as external knowledge. Despite empirically validated effective, mainstream retrieval strategies (lexical-based, semantic-based, or both) rely solely on surface-level code similarity, which can lead to mismatches between the retrieved examples and the actual repair needs \cite{Vulkey}. Instead, APPatch \cite{appatch} incorporates historical exemplars that closely match the root cause of the testing sample, as well as their corresponding fixing strategies and ground-truth patches, into the prompt for patch generation. Nevertheless, with the expansion of exemplar scale, such a purely LLM-driven selection process is time-consuming and cost-expensive. In addition, even identified the root cause of the vulnerability under the guidance of developer-provided repair oracle (typically unavailable in practice), APPatch fails to produce correct patches as LLMs may be misled by exemplars with similar root causes but completely different repair strategies. As demonstrated in Fig. \ref{fig:motivation} (right), one of the selected exemplars CVE-2018-16419 \cite{CVE-2018-16419} contains a stack-based buffer overflow vulnerability in which the variable \code{bufsize} is overwritten with \code{file->size} retrieved from the smartcard. If \code{file->size} is larger than 2048 bytes, the read operation will write data beyond the bounds of the fixed stack buffer. This vulnerability was fixed officially by capping the read buffer size to the smaller of the untrusted file size and the fixed 2048-byte capacity of the stack buffer. Unfortunately, the patch generated by APPatch is neither (\ding{182}) compilable due to the use of undefined macro \code{AF\_ARRAY\_SIZE} nor (\ding{183}) complete as \code{numCoefficients} should be at least 7.

\subsection{Key Ideas}
Based on the above observations, we propose a knowledge-enhanced agentic AVR approach that jointly addresses root cause analysis accuracy and demonstration quality.

\textbf{(1) Fact-grounded agentic vulnerability diagnosis.}
Instead of relying on purely static LLM reasoning or dynamic localization tools, we introduce a tool-augmented diagnostic agent that performs ReAct-style \cite{yao2023react} reasoning on vulnerable code to collect verified program facts for reliable root cause analysis. These grounded facts facilitate minimizing inherent hallucinations of LLMs \cite{DBLP:journals/corr/abs-2410-09997} in subsequent patch generation.

\textbf{(2) Knowledge-level retrieval-augmented patch generation.}
Prior approaches suffer from suboptimal retrieval strategies, where root cause-similar exemplars with divergent repair patterns easily mislead LLMs into generating invalid patches. To effectively and efficiently reuse prior fix knowledge to unleash the potential of LLMs for patching, we decompose the single similarity metric into dual complementary views \cite{view}, including control-flow and data-flow, with dedicated knowledge bases for retrieval-augmented patch generation.

\section{Methodology}\label{Methodology}

\begin{figure}[t]
  \centering
  \includegraphics[width=\linewidth]{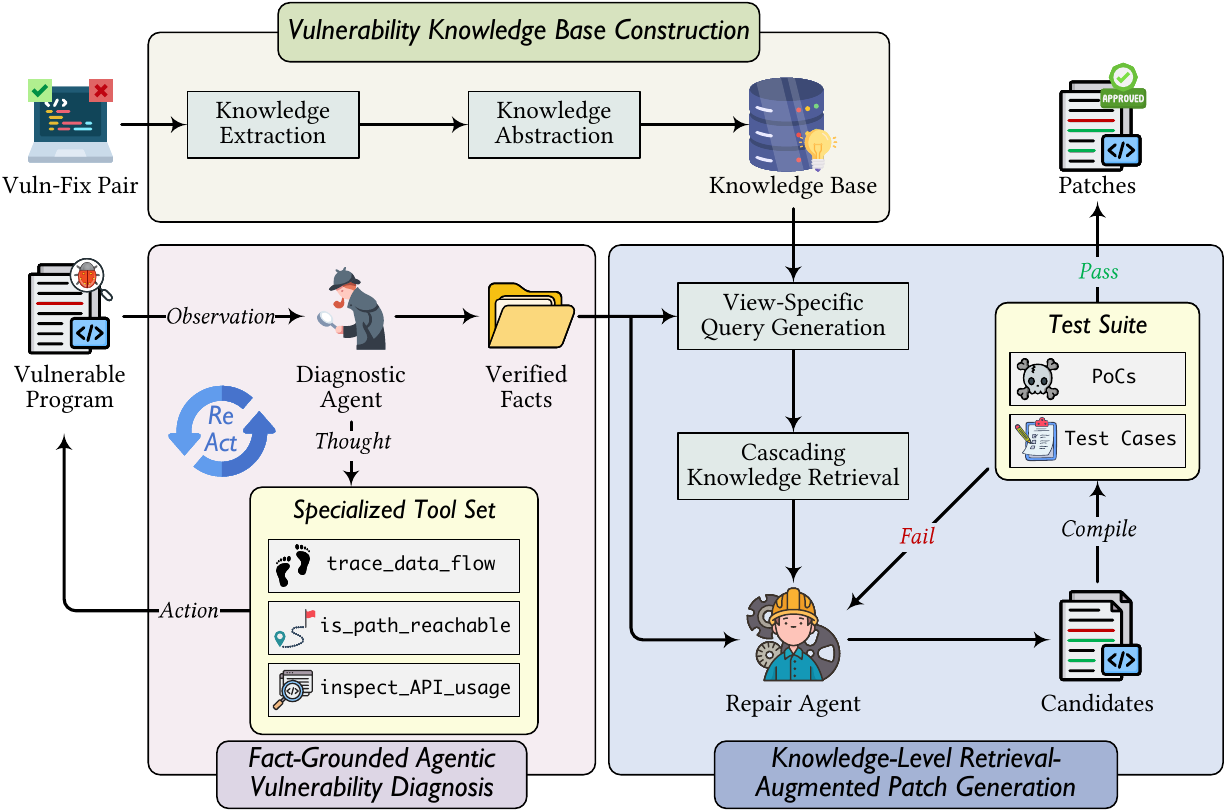}
\caption{Overview of \sysname.}
\label{fig:Overview}
\vspace{-5mm}
\end{figure}

\subsection{Overview}\label{Overview}
Fig.~\ref{fig:Overview} illustrates the overall workflow of \sysname, which operates through three main phases: vulnerability knowledge base construction, fact-grounded agentic vulnerability diagnosis, and knowledge-level retrieval-augmented patch generation.

\begin{itemize}[leftmargin=1em]
  \item \textbf{Vulnerability Knowledge Base Construction:}
  Given a group of historical vulnerability-patch pairs, \sysname instructs the LLM to distill view-specific multi-dimensional vulnerability knowledge under the supervision of ground-truth patches. All the knowledge items are then aggregated to form the final knowledge base for retrieval.

  \item \textbf{Fact-Grounded Agentic Vulnerability Diagnosis:}
  For a to-be-patched vulnerability, the diagnostic agent equipped with crafted static analysis tools conducts iterative, demand-driven reasoning and exploration to collect verified facts related to suspicious program points.

  \item \textbf{Knowledge-Level Retrieval-Augmented Patch Generation:}
  Based on the diagnostic report, \sysname independently retrieves relevant historical samples from the constructed knowledge base in accordance with each view-specific semantic query, and incorporates them along with the distilled vulnerability knowledge into the prompt to synthesize a patch. The patch will be validated against a series of automated checks in a closed feedback loop.
  
\end{itemize}

% \begin{algorithm}[t]
% \caption{Offline Index Construction}
% \label{alg:offline_index}
% \textbf{Input:} $\mathcal{D} = \{(b_i, p_i)\}_{i=1}^{N}$: dataset of (buggy code, patch) pairs \\
% \textbf{Output:} $\mathcal{I} = \{\mathcal{I}_{df}, \mathcal{I}_{cf}, \mathcal{I}_{api}\}$: three-view FAISS indices
% \begin{algorithmic}[1]
% \Function{BuildIndex}{$\mathcal{D}$}
%     \For{each $(b_i, p_i) \in \mathcal{D}$}
%         \For{each view $v \in \{df, cf, api\}$}
%             \If{$p_i \neq \emptyset$}
%                 \State $\mathbf{R}_i[v] \leftarrow$ \Call{PatchGuidedGen}{$b_i,\ p_i,\ v$}
%                     \Comment{// Patch as directional oracle; no diff code in output}
%             \Else
%                 \State $\mathbf{R}_i[v] \leftarrow$ \Call{BlindGen}{$b_i,\ v$}
%                     \Comment{// Fallback: patch unavailable}
%             \EndIf
%             \State $r_v \leftarrow$ \Call{Distill}{$\mathbf{R}_i[v],\ v$}
%                 \Comment{// Compress to unified JSON schema; strip redundancy}
%             \State $\mathbf{e}_v \leftarrow \mathcal{E}(r_v)$
%                 \Comment{// Encode distilled view to embedding vector}
%             \State $\mathcal{I}_v$.\Call{Add}{$\mathbf{e}_v,\ \text{meta}(b_i, p_i, r_v, v)$}
%                 \Comment{// Insert into view-specific FAISS index}
%         \EndFor
%     \EndFor
%     \State \Return $\mathcal{I}$
% \EndFunction
% \end{algorithmic}
% \end{algorithm}

\subsection{Vulnerability Knowledge Base Construction}\label{Offline Index Construction}
% As mentioned above, retrieving relevant exemplars at the instance- or root cause-level may degrade the patch generation performance as project-specific contexts induce superficial mimicry, while misaligned repair strategies can mislead LLMs. 
Inspired by how developers understand vulnerabilities \cite{DBLP:conf/ccs/ZhaoZ0ZZY20}, \sysname distills multi-dimensional knowledge from historical vulnerabilities and fixes to better guide LLMs. To achieve this, we need to address three tasks: (\ding{182}) Knowledge Extraction, (\ding{183}) Knowledge Abstraction, and (\ding{184}) Knowledge Storage.

\subsubsection{Knowledge Extraction}
Different from APPatch \cite{appatch}, which characterizes vulnerabilities through single-view root cause analysis, we independently represent and extract vulnerability knowledge from dual views, including data-flow and control-flow, which have been proven to be essential for understanding vulnerability semantics \cite{view,MVD}. For each view, we instruct LLMs to distill the following five knowledge items from three dimensions with dedicated prompts.

\noindent\textbf{Root Cause.}
This dimension describes the fundamental security issue responsible for the violation of security by comparing the vulnerable code and its corresponding patch.

\noindent\textbf{Vulnerable Behaviors.}
This dimension concludes the unexpected program behaviors that raise vulnerabilities. For instance, a memory leak occurs when allocated memory is not freed along some program path. The behavior can be described from three perspectives: (\ding{182}) \emph{Sensitive APIs} are basic operations (e.g., \code{malloc}) that vulnerabilities typically originate from their improper usage; (\ding{183}) \emph{Taint Flow} describes how an untrusted input propagate from an entry point to the security-sensitive sink; and (\ding{184}) \emph{Control Branch} records the specific control conditions and precise path information, illustrating under what conditions the vulnerabilities are triggered.

\noindent\textbf{Fixing Strategies.}
This dimension summarizes the fixing solution by analyzing the vulnerability-patch pair.

\begin{tcolorbox}[breakable,colback=white,colframe=black,boxrule=1pt,
         left=2pt,right=2pt,top=2pt,bottom=2pt,
        title=Prompt Template for Knowledge Extraction]

\textbf{[System Prompt]}

You are a senior security expert experienced in vulnerability diagnosis. Your objective is to meticulously analyze the given vulnerability-patch pair from the view of \emph{data (/control) flow} to extract high-level vulnerability knowledge. 

\textbf{[User Prompt]}

You are given the following information for analysis:
\begin{itemize}[leftmargin=1em]
    \item \textbf{CVE Description:} \{\code{CVE\_description}\}
    \item \textbf{Vulnerable Function:} \{\code{vuln\_func}\}
    \item \textbf{Patched Function:} \{\code{patch\_func}\}
\end{itemize}

Please perform your analysis by strictly following the hierarchical reasoning process below. \textbf{Think step-by-step}:

\begin{itemize}[leftmargin=1em]
    \item \textbf{Step 1: Code Change Comprehension.}
    What was added, deleted, or modified to edit the vulnerable function into the patched version.

    \item \textbf{Step 2: Vulnerable Behavior Reasoning.}
    Which \emph{security-sensitive APIs} are used in the vulnerable function, and how they can be called to perform dangerous operations via a specific \emph{taint flow (/control branch)}.

    \item \textbf{Step 3: Pairwise Analysis.}
    Based on the above analysis, summarize the generalizable \emph{root cause} that leads to the vulnerability and the specific \emph{solution to fix it}.
\end{itemize}

Please return the results in a structured JSON format.
\end{tcolorbox}

\noindent\textbf{Prompt Design.}
The prompt template for extracting view-specific knowledge is shown above. We start from a system prompt to inform the model of the role it plays. Then it is followed by the data we present to the LLM, which consists of the CVE description and vulnerability-patch pair. The subsequent segment outlines the task directives provided to the LLM. In particular, \sysname employs a hierarchical reasoning strategy \cite{DBLP:conf/nips/Wei0SBIXCLZ22}, which gradually progresses from observing an editing action to understanding \textbf{why} the vulnerability occurs and \textbf{how} it is fixed, mimicking an expert's cognitive workflow in vulnerability diagnosis. Finally, the LLM is mandated to output a structured report formatted in JSON that contains \code{view\_type}, \code{root\_cause}, \code{risky\_calls}, \code{taint\_flow} (data-flow-only), \code{control\_branch} (control-flow-only), and \code{fixing\_strategy}.

\subsubsection{Knowledge Abstraction}
As different vulnerability instances might share high-level commonality (e.g., the similar root causes and fixing strategies), \sysname further performs abstraction to distill more general knowledge that is less bound to concrete code implementation details. Following Du et al. \cite{vul-rag}, we instruct LLMs to abstract the concrete code elements, including method invocations, variable names, and types, in the extracted vulnerability causes and fixing solutions.

\subsubsection{Knowledge Storage}\label{ks}
For each vulnerability instance, its dual-view vulnerability knowledge is stored in separate vector databases for indexing and retrieval. Specifically, \sysname leverages \code{all-MiniLM-L6-v2}\footnote{\url{https://huggingface.co/sentence- transformers/all-MiniLM-L6-v2}}, a lightweight sentence embedding model, to encode the root cause into a 384-dimensional feature vector, while retaining the other knowledge items and the corresponding vulnerability-patch pair as metadata. To enable efficient retrieval in high-dimensional spaces, we use Faiss \cite{DBLP:journals/tbd/DouzeGDJSMLHJ26} as our search engine.

\subsection{Fact-Grounded Agentic Vulnerability Diagnosis}\label{Agentic-Diagnosis}
The online repair stage begins with an in-depth diagnosis of the input vulnerable code, aiming to understand the root cause and characteristics of the vulnerability. To this end, we design a tool-augmented diagnostic agent following the ReAct paradigm \cite{yao2023react}. This interleaved thought-action-observation cycle enables demand-driven exploration of the repository, with observations and intermediate reasoning accumulated across turns. Especially, instead of relying on general-purpose operation interfaces \cite{patchagent,DBLP:conf/nips/YangJWLYNP24} or syntax-based retrieval \cite{san2patch,DBLP:conf/issta/0002RFR24}, our diagnostic agent is equipped with three specialized static analysis tools to collect semantic facts related to suspicious program points:

\begin{itemize}[leftmargin=1em]
    \item \code{trace\_data\_flow} requires the LLM to specify a variable $v$ of interest at statement $l$, and returns its intra-procedural taint propagation paths via backward traversal on the pre-computed def-use chains. It allows to verify whether untrusted, unchecked, or otherwise unsafe values can influence the vulnerable operation.

    \item \code{is\_path\_reachable} requires the LLM to specify a source statement $s$, a target statement $t$, and a variable $v$ of interest, and returns the intra-procedural control-flow paths from $s$ to $t$. For each path, the tool reports the involved statements, branch predicates, and dominating conditions related to $v$. It allows to verify whether a suspicious operation is executable under particular control-flow conditions and whether relevant checks are encountered before the target statement.

    \item \code{inspect\_api\_usage} requires the LLM to specify a sensitive API call in the vulnerable code, and returns the syntactic usage context of the API, including the callee, receiver object, arguments, return value, enclosing function, and surrounding statements. If the callee is locally defined in the same file, the tool further returns its implementation slice. These observations help the agent understand the actual behavior of project-specific APIs from code evidence.
\end{itemize}

We utilize the \code{tree-sitter} \cite{tree-sitter} parsing library to provide the above primitive tools. The ReAct loop will not stop until the maximum interaction turns $K$ (where $K=15$ in our experiments) reaches or sufficient program facts are received. These facts are then organized as structured diagnostic report, which forms the basis of the subsequent patch generation.

\subsection{Knowledge-Level Retrieval-Augmented Patch Generation}\label{Online-Retrieval}
Given the structured diagnostic report, \sysname synthesizes dual-view semantic queries, and retrieves relevant historical exemplars from the view-specific knowledge base in parallel for patch generation. Only patches that pass all quality criteria are selected as final solutions; otherwise, they are iteratively refined with execution feedback. 

\subsubsection{View-Specific Query Generation}
Unlike existing retrieval-augmented AVR approaches that solely use vulnerable code \cite{FixRepair,yin2024thinkrepair} as the query, \sysname prompts LLMs to generate view-specific root causes based on the diagnostic report to find exemplars that share high-level commonality with the given vulnerability. It's noteworthy that, despite the similarity to APPatch \cite{appatch} in terms of the design idea, our vulnerability cause is derived from the verified program facts instead of free-form LLM reasoning, which suffers from the hallucination issue and shows unstable bias with different prompts \cite{vul-rag}.

\subsubsection{Cascading Knowledge Retrieval}
To select relevant historical exemplars and their corresponding knowledge items (e.g., referable fixing strategies) as demonstrations, \sysname adopts a cascading retrieval pipeline. Initially, \sysname retrieves the Top-10 candidates separately from the view-specific vulnerability knowledge base. We use the same embedding model \code{all-MiniLM-L6-v2} to encode dual-view root causes (i.e., query), and compute the squared L2 (Euclidean) distance between the query embedding and the key embeddings stored in vulnerability knowledge base. A shorter distance indicates greater similarity. However, retrieving the exemplar with the most similar root cause does not guarantee finding the one that helps fix the target vulnerability as they may correspond to distinct repair patterns \cite{Vulkey}. To address this, we further re-rank the retrieved exemplars using a hybrid re-ranker.

Given a set of retrieved candidates, the hybrid re-ranker measures the similarity between the target vulnerability and historical exemplars at the code- and behavior-level. 

\noindent\textbf{Code Similarity.}
Unlike previous studies training the embedding model from scratch \cite{FixRepair}, we directly employ the pre-trained CodeLlama \cite{DBLP:journals/corr/abs-2308-12950} to encode the vulnerable code $q^{\text{code}}$ to be patched and the vulnerable code $d_i^{\text{code}} \in \mathcal{D}$ of candidate exemplar $d_i$. \sysname then computes the cosine similarity $Cos(\boldsymbol{q}^{\text{code}},\boldsymbol{d}_i^{\text{code}})$ between their embeddings to measure their semantic relevance.

\noindent\textbf{Behavior Similarity.}
We utilize BM25 \cite{DBLP:journals/ipm/SaltonB88} to retrieve exemplars with similar vulnerable behaviors. To be specific, \sysname treats the structured diagnostic report $q^{\text{beh}}$ as a token sequence, and converts it into a bag-of-words representation. Similarly, three knowledge items, including sensitive APIs, taint flow, and/or control branch) of each candidate exemplar $d_i$ are concatenated as $d_i^{\text{beh}}$. It then calculates the lexical similarity $BM25(q^{\text{beh}},d_i^{\text{beh}})$ of the target vulnerability and candidate exemplars at the behavior-level.

\noindent\textbf{Hybrid Re-ranking.}
To take both code- and behavior-level information into account, we adopt a hybrid approach that integrates CodeLlama and BM25 in a unified manner. The hybrid similarity score is calculated as follows:
\begin{equation}
    f_\phi(q,d_i) =
    \lambda Cos(q^{\text{code}}, d_i^{\text{code}}) +
    (1-\lambda) \text{BM25}(q^{\text{beh}}, d_i^{\text{beh}})
\end{equation}
where $\lambda$ is a weight coefficient to balance the two re-rankers. Note that the BM25 score is normalized between 0-1 to ensure that it is on the same scale as cosine similarity.

Adhering to the best practice \cite{DBLP:conf/kbse/GaoWGWZL23}, we select the Top-4 relevant exemplars based on this combined similarity score (placed in ascending order) to guide the patch generator.

\subsubsection{Patching and Validation}
Based on the (\ding{182}) diagnostic report, (\ding{183}) retrieved exemplars, and (\ding{184}) their corresponding knowledge items, the repair agent performs in-context learning over historical fixing strategies to synthesize an initial patch, formatted as a multi-hunk unified diff. As LLMs frequently hallucinate line numbers or context lines, we applies a deterministic patch correction algorithm \cite{patchagent}, which automatically adjusts the diff using a minimal edit distance heuristic.

\noindent\textbf{Patch Refinement with Feedback.}
Once a valid patch is output, \sysname recompiles the target project and applies two progressively stricter validation checks in turn: replaying the triggering PoC, and re-executing the test suite. If any check fails, \sysname returns the failure output to the repair agent to guide its next exploration or edit. This closed feedback loop continues until all checks are passed or a pre-defined iteration limit $N$ is reached.

\section{Evaluation}\label{Evaluation}

\subsection{Research Questions}\label{questions}

Our work seeks to answer four Research Questions (RQs):

\begin{itemize}[leftmargin=1em]
    \item \textbf{RQ1:} How does \sysname perform in generating patches for real-world vulnerabilities?
    \begin{itemize}
    \item \textbf{RQ1a:} Which LLM backbone is the best for \sysname?
    \item \textbf{RQ1b:} What about the inference cost of \sysname?
    \end{itemize}    
    \item \textbf{RQ2:} How effective is \sysname compared to baselines?
    \item \textbf{RQ3:} How do various components affect the overall performance of \sysname?
    \item \textbf{RQ4:} What is the influence of hyper-parameters on the performance of \sysname?
\end{itemize}

\subsection{Baselines}\label{baselines}
We evaluate \sysname against four groups of baselines, including constraint-based \cite{extractfix,vulnfix}, learning-based \cite{vulmaster,vulrepair}, LLM-based \cite{appatch,looprepair}, and agent-based approaches \cite{patchagent,san2patch}.

\begin{itemize}[leftmargin=1em]   
    \item \textbf{Constraint-based approaches} employ static analysis or dynamic execution of test cases or exploits to generate candidate patches. ExtractFix~\cite{extractfix} derives crash-free constraints via sanitizer-defined rules, while VulnFix~\cite{vulnfix} infers invariants by mutating program states to guide the generation process.
    
    \item \textbf{Learning-based approaches} fine-tune pre-trained code models to generate the fixed version of a vulnerable function \cite{vulrepair}. The state-of-the-art VulMaster \cite{vulmaster} integrates diverse information, encompassing vulnerable code structures and expert knowledge to achieve a high repair capability. 
    
    \item \textbf{LLM-based approaches} use carefully crafted prompts to guide LLMs in patch generation. LoopRepair~\cite{looprepair} instructs LLMs to generate candidate patches based on predicted fix locations, and designs a  taint trace-guided ranking strategy to select the optimal patch during iteration. APPatch \cite{appatch} combines semantics-aware slicing, adaptive prompting, and exemplar selection to improve patch generation.

    \item \textbf{Agent-based approaches} embed the LLM in an iterative loop equipped with tools for code navigation, compilation, and test execution. Starting from the sanitizer report, PatchAgent \cite{patchagent} and San2Patch \cite{san2patch} retrieve context iteratively and reasons about the vulnerability statically.
\end{itemize}

\subsection{Dataset}\label{datasets}
We evaluate \sysname using the \textsc{Vul4C} \cite{DBLP:conf/uss/Hu0SGZXY025} dataset, which contains 68 real-world C/C++ vulnerabilities, each paired with the associated PoCs and test cases. 13 vulnerabilities are excluded as they could not be processed by constraint- or learning-based baselines, resulting in a final evaluation set of \textbf{55} CVEs, which correspond to 10 software products and six vulnerability types, ranging from buffer overflows and integer overflows to more complex use-after-free vulnerabilities.

\subsection{Evaluation Metrics}\label{metrics}
In line with prior work \cite{san2patch,patchagent,DBLP:journals/corr/abs-2511-11019}, we adopt two widely used metrics for evaluating approaches:

\begin{itemize}[leftmargin=1em]   
    \item \textbf{Repair Rate} measures the overall repair performance. Let $a$ denotes the number of vulnerabilities for which an AVR tool generates at least one compilable patch that successfully pass security tests (via PoCs) and functionality tests (via test cases), and $b$ the total number of vulnerabilities under repair. This metric is defined as  $a/b \times 100\%$.
    \item \textbf{Average Cost} calculates the average LLM API cost per instance from token usage, regardless of success or failure.
\end{itemize}

\subsection{Implementation}
\noindent\textbf{Experiment Environment.}
All experiments were conducted on an Ubuntu 22.04.5 LTS server with an AMD Ryzen 7 5700G @3.80GHz, 128GB RAM, and one NVIDIA A100 GPU (80GB). Throughout the experiments, LLM APIs were uniformly accessed via OpenRouter\footnote{\url{https://openrouter.ai}}. Unless otherwise stated, the default LLM backbone used was Google's Gemini-3.1-Pro-Preview as it struck a good trade-off between effectiveness and efficiency (\S\ref{RQ1-Results}). In light of the best performance (\S\ref{sens}), we report experimental results in a setting with the weight coefficient $\lambda$ as 0.3.

\noindent\textbf{Knowledge Preparation.}
To collect exemplars for retrieval, we employ \textsc{PrimeVul} \cite{DBLP:conf/icse/DingFISCAWR025}, a high-quality function-level vulnerability dataset that addresses label inaccuracies through combined automated labeling and manual verification, coupled with rigorous de-duplication. \textsc{PrimeVul} features 5,480 vulnerability-patch pairs across 130 CWE categories, sourced from 754 open-source projects. Given the demand for rapid response and cost-effectiveness in generating a large volume of vulnerability-related knowledge items, we adopt GPT-5-mini-2025-08-07 during the offline knowledge base construction.

\noindent\textbf{Patch Validation.}
To ensure reproducibility, each vulnerability is isolated within a Docker container and includes an automated script that compiles the target first, and then executes the PoC and unit tests to confirm whether it is successfully fixed. Adhering to the best practice \cite{yin2024thinkrepair}, we adopt five times of interaction iterations (i.e., $N=5$) as the default setting.

\begin{table}[t]
\centering
\caption{Patching results of \sysname across various vulnerability types under different LLMs.}
\label{tab:rq1}
% \setlength{\tabcolsep}{4pt}
% \small
\begin{threeparttable}
\renewcommand\arraystretch{0.8}
\resizebox{\columnwidth}{!} {
\begin{tabular}{clccccccc}
\toprule
\multirow{2.5}{*}{\textbf{Arch.}} & \multirow{2.5}{*}{\textbf{Backbone Model}} &
\multicolumn{6}{c}{\textbf{Vulnerability Type\textsuperscript{\dag}}} & \multirow{2.5}{*}{\textbf{Total}}\\
\cmidrule(lr){3-8}
~ & ~ & \textbf{BO} & \textbf{IO} & \textbf{DZ} & \textbf{UAF} & \textbf{NPD} & \textbf{IL} & ~\\
\midrule
\multirow{5}*{\rotatebox{90}{Standalone}} & GPT-5.1            & 18/38 & 3/6 & 2/5 & 1/2 & 0/2 & 0/2 & 24/55 \\
~ & Claude-Sonnet-4-6  & 21/38 & 3/6 & 2/5 & 2/2 & 1/2 & 1/2 & 30/55 \\
~ & Gemini-3.1-Pro     & 16/38 & 5/6 & 2/5 & 0/2 & 0/2 & 1/2 & 24/55 \\
~ & Qwen3.5-Plus       & 19/38 & 3/6 & 1/5 & 2/2 & 1/2 & 1/2 & 27/55 \\
~ & DeepSeek-V3.2      & 16/38 & 5/6 & 1/5 & 1/2 & 0/2 & 1/2 & 24/55 \\
\midrule
\multirow{5}*{\rotatebox{90}{\sysname}} & GPT-5.1            & 30/38 & 6/6 & 4/5 & 2/2 & 2/2 & 2/2 & 46/55 \\
~ & Claude-Sonnet-4-6  & 30/38 & 6/6 & 3/5 & 1/2 & 2/2 & 2/2 & 44/55 \\
~ & Gemini-3.1-Pro     & 30/38 & 6/6 & 4/5 & 2/2 & 2/2 & 2/2 & 46/55 \\
~ & Qwen3.5-Plus       & 28/38 & 6/6 & 3/5 & 1/2 & 2/2 & 1/2 & 41/55 \\
~ & DeepSeek-V3.2      & 29/38 & 4/6 & 2/5 & 1/2 & 2/2 & 2/2 & 40/55 \\
\bottomrule
\end{tabular}}
  \begin{tablenotes}
        \scriptsize
        \item[\dag] BO: Buffer Overflow, IO: Integer Overflow, DZ: Divide by Zero, UAF: Use-After- \\ Free, NPD: Null Pointer Dereference, IL: Infinite Loop.
  \end{tablenotes}
\end{threeparttable}
\vspace{-4mm}
\end{table}

\section{Experimental Results}\label{Experimental Results}

\subsection{RQ1: Effectiveness of \sysname}\label{RQ1-Results}
\subsubsection{[RQ1a]  Optimal LLM Selection}\

\noindent\textbf{Experiment Setup.}
In order to analyze how different backbone models affect the performance of \sysname, we select a range of frontier LLMs, including GPT-5.1-2025-11-13, Claude-Sonnet-4-6, Gemini-3.1-Pro-Preview, Qwen3.5-Plus-2026-04-20, and DeepSeek-V3.2-0324, as the backend. Apart from being integrated into our approach, they are also naively prompted under the zero-shot setting \cite{DBLP:conf/sp/PearceTAKD23} to generate fixes at the place in the code where the original developers patched each vulnerability to act as the control group. To obtain deterministic and reproducible results, we set the temperature parameter to 0 for LLMs that support this option, while retaining default configurations for those that do not.

\noindent\textbf{Results.}
Table~\ref{tab:rq1} reports the patching results for each model across different vulnerability types. Overall, \sysname delivers strong performance across all types by leveraging diverse LLMs, and consistently outperform standalone LLMs. We can observe that, when paired with GPT-5.1 (or Gemini-3.1-Pro), \sysname stands out with the highest repair rate, successfully addressing 46 out of 55 (83.64\%) vulnerabilities, excelling particularly in integer overflow, use-after-free, null pointer dereference, and infinite loop with a perfect 100\% success rate. For buffer overflow and divide by zero, the success rates are slightly lower, at 78.95\% (30/38) and 80.00\% (4/5), respectively. Claude-Sonnet-4-6 also performs impressively, achieving an 80.00\% (44/55) repair rate. In contrast, two open-sourced LLMs Qwen3.5-Plus and DeepSeek-V3.2 lag behind, patching 41 and 40 vulnerabilities, respectively. Despite that, they still outperform the best-performing standalone Claude-Sonnet-4-6, which patches only 30 vulnerabilities. Additionally, we find that the performance gains brought by incorporating \sysname vary for different LLMs. For example, when paired with \sysname, the number of vulnerabilities patched by GPT-5.1 increases from 24 to 46, with the most significant improvement of 91.67\%. A similar situation can be observed as in Gemini-3.1-Pro. These results underscore the superiority of \sysname in effectively patching vulnerabilities.

\rqbox{\ding{45} \textbf{Answer to RQ1a}
    $\blacktriangleright$ \sysname delivers strong performance across all vulnerabilities types by leveraging diverse LLMs. Among the tested models, GPT-5.1 and Gemini-3.1-Pro achieve the highest repair performance, successfully addressing 46 out of 55 vulnerabilities. $\blacktriangleleft$}

\begin{table}[!t]
\centering
\caption{Cost analysis of \sysname under different LLMs.}
\label{tab:rq5}
\renewcommand\arraystretch{0.8}
\resizebox{\columnwidth}{!} {
\begin{tabular}{lccccccc}
\toprule
\multirow{2.5}{*}{\textbf{Model}} &
\multirow{2.5}{*}{\makecell[c]{\textbf{Repair} \\ \textbf{Rate (\%)}}} &
\multirow{2.5}{*}{\makecell[c]{\textbf{Avg.} \\ \textbf{Iter.}}} &
\multicolumn{4}{c}{\textbf{Time Overhead (s)}} &
\multirow{2.5}{*}{\makecell[c]{\textbf{Avg.} \\ \textbf{Cost (\$)}}} \\
\cmidrule(lr){4-7}
& & & \textbf{Min} & \textbf{Max} & \textbf{Med} & \textbf{Avg.} & \\
\midrule
GPT-5.1           & 83.64 & 1.70 & 8 & 398 & 25 & 61.24 & 1.15 \\
Claude-Sonnet-4-6 & 80.00 & 1.89 & 7 & 529 & 27 & 68.80 & 1.38 \\
Gemini-3.1-Pro    & 83.64 & 1.20 & 5 & 409 & 22 & 49.43 & 0.30 \\
Qwen3.5-Plus      & 74.55 & 1.54 & 9 & 527 & 22 & 62.15 & 0.12 \\
DeepSeek-V3.2     & 72.73 & 1.60 & 7 & 507 & 32 & 75.60 & 0.09 \\
\bottomrule
\end{tabular}}
\vspace{-4mm}
\end{table}

\subsubsection{[RQ1b] Efficiency}\

\noindent\textbf{Experiment Setup.}
To assess the practical feasibility of \sysname in real-world scenarios, we further measure the average time (in seconds) and API costs (in USD) required per vulnerability for each LLM to generate patches. 

\noindent\textbf{Results.}
The statistics are summarized in Table \ref{tab:rq5}. Overall, Gemini-3.1-Pro, which achieves the highest repair success rate at 83.64\%, strikes the optimal cost-performance trade-off. Particularly, the average time required per vulnerability is 49.43 seconds, with an approximate cost of \$0.30, demonstrating that it can be reasonably applied in real-world environments \cite{DBLP:conf/icse/NollerS0R22}. In contrast, despite the comparable effectiveness, GPT-5.1 and Claude-Sonnet-4-6 incur a higher cost (\$1.15/\&1.38) and time overhead (61.24/68.80 seconds), respectively. DeepSeek-V3.2 offers the most cost-effective solution at \$0.09 per vulnerability, though with a lower repair rate of 72.73\%. Qwen3.5-Plus falls in the middle range, repairing 74.55\% of vulnerabilities with an average of 1.54 iterations, resulting in a lower time overhead (62.15 seconds) and cost (\$0.12) per vulnerability.

\rqbox{\ding{45} \textbf{Answer to RQ1b}
    $\blacktriangleright$ Gemini-3.1-Pro achieves the optimal cost-benefit trade-off, repairing 83.64\% of vulnerabilities at a lower cost (\$0.30) and within practical timeframes (under 50 seconds) per vulnerability. $\blacktriangleleft$}

\subsection{RQ2: Comparison with State-of-the-Art Work}\label{RQ2:Results}
\begin{table}[t]
\centering
\caption{Comparison with state-of-the-art AVR approaches.}
\label{tab:rq2}
\renewcommand\arraystretch{0.8}
\setlength{\tabcolsep}{0.4\tabcolsep}
\resizebox{\columnwidth}{!} {
\begin{tabular}{lll|cc|cc|cc|cc|c}
\toprule\textbf{Project} & \textbf{CVE-ID} & \textbf{Type} & \textbf{V.R.} &
\textbf{V.M.} & \textbf{V.F.} & \textbf{E.F.} &
\textbf{L.R.} & \textbf{AP.} &
\textbf{P.A.} & \textbf{S.P.} &
\textbf{K.R.} \\
\midrule
\multirow{17}*{libtiff} & CVE-2006-2025  & IO  & \XM & \XM & \XM & \XM & \CM & \XM & \CM & \CM& \CM \\
~ & CVE-2010-2481  & BO  & \XM & \XM & \XM & \XM & \XM & \CM & \XM & \CM & \CM \\
~ & CVE-2013-4243  & BO  & \XM & \XM & \CM & \XM & \CM & \CM & \CM & \CM & \CM \\
~ & CVE-2016-10092 & BO  & \CM & \XM & \XM & \XM & \CM & \XM & \XM & \XM & \XM \\
~ & CVE-2016-10093 & BO  & \XM & \XM & \XM & \XM & \CM & \XM & \CM & \CM & \CM \\
~ & CVE-2016-10094 & IO  & \XM & \XM & \XM & \XM & \CM & \XM & \CM & \CM & \CM \\
~ & CVE-2016-10266 & DZ  & \XM & \XM & \CM & \XM & \CM & \CM & \CM & \CM & \CM \\
~ & CVE-2016-10267 & DZ  & \XM & \XM & \XM & \XM & \CM & \CM & \CM & \CM & \CM \\
~ & CVE-2016-10268 & IO  & \XM & \XM & \XM & \XM & \CM & \CM & \CM & \CM & \CM \\
~ & CVE-2016-10271 & BO  & \XM & \XM & \XM & \XM & \CM & \XM  & \XM & \CM & \CM\\
~ & CVE-2016-10272 & BO  & \XM & \XM & \XM & \XM & \CM & \XM & \CM & \XM & \CM \\
~ & CVE-2016-5321  & BO  & \XM & \CM & \CM & \CM & \CM & \XM & \CM & \CM & \CM\\
~ & CVE-2017-5225  & BO  & \XM & \XM & \XM & \XM & \CM & \CM & \CM & \XM & \CM \\
~ & CVE-2017-7595  & DZ  & \XM  & \XM & \CM & \XM & \CM & \XM & \CM & \CM & \CM\\
~ & CVE-2017-7598  & DZ  & \XM  & \XM & \XM & \XM & \CM & \CM & \XM & \XM & \CM \\
~ & CVE-2017-7601  & BO  & \XM  & \XM & \CM & \XM & \CM & \CM & \CM & \CM & \CM \\
~ & CVE-2017-7602  & IO  & \XM  & \XM & \XM & \XM & \XM & \CM & \CM & \XM & \CM \\
\midrule
\multirow{3}*{libxml2} & CVE-2012-5134  & BO  & \XM & \XM & \CM & \XM & \CM & \XM & \CM & \CM & \CM \\
~ & CVE-2016-1838  & BO  & \XM & \XM & \XM & \XM & \XM & \XM & \CM & \XM & \XM \\
~ & CVE-2017-5969  & NPD & \XM & \XM & \CM & \XM & \XM & \XM & \XM & \CM & \CM \\
\midrule
\multirow{3}*{libarchive} & CVE-2016-10349 & BO & \XM & \XM & \XM & \XM & \CM & \CM & \CM & \CM & \CM \\
~ & CVE-2016-10350 & BO & \XM & \XM & \XM & \XM & \CM & \CM & \CM & \CM & \CM \\
~ & CVE-2016-5844  & IO & \CM & \CM & \CM & \XM & \CM & \CM & \CM & \CM & \CM \\
\midrule
\multirow{2}*{libzip} & CVE-2017-12858 & UAF & \XM & \XM & \XM & \XM & \CM & \XM & \CM & \CM & \CM \\
~ & CVE-2017-14107 & BO  & \XM & \XM & \XM & \XM & \CM & \XM & \CM & \XM & \CM \\
\midrule
\multirow{3}*{graphicsmagick} & CVE-2017-12937 & BO  & \XM & \XM & \XM & \XM & \CM & \CM & \CM & \CM & \CM\\
~ & CVE-2017-14103 & UAF & \CM & \XM & \CM & \XM & \CM & \XM & \CM & \CM & \CM\\
~ & CVE-2017-14649 & NPD & \XM & \XM & \XM & \XM & \XM & \XM & \XM & \CM & \CM \\
\midrule
\multirow{11}*{audiofile} & CVE-2017-6827 & BO & \XM & \XM & \XM & \XM & \XM & \CM & \CM & \XM & \CM \\
~ & CVE-2017-6828 & BO & \XM & \XM & \XM & \XM & \XM & \XM & \CM & \CM & \CM \\
~ & CVE-2017-6829 & BO & \XM & \XM & \XM & \XM & \CM & \CM & \CM & \CM & \CM \\
~ & CVE-2017-6830 & BO & \CM & \XM & \CM & \XM & \CM & \XM & \CM & \CM & \CM \\
~ & CVE-2017-6831 & BO & \XM & \XM & \XM & \XM & \CM & \CM & \XM & \CM & \CM \\
~ & CVE-2017-6832 & BO & \XM & \XM & \XM & \XM & \CM & \CM & \CM & \CM & \CM \\
~ & CVE-2017-6833 & BO & \XM & \XM & \XM & \XM & \CM & \CM & \CM & \XM & \XM \\
~ & CVE-2017-6834 & BO & \XM & \XM & \XM & \XM & \CM & \XM & \XM & \CM & \CM \\
~ & CVE-2017-6835 & BO & \XM & \XM & \XM & \XM & \XM & \CM & \CM & \CM & \CM \\
~ & CVE-2017-6836 & BO & \XM & \XM & \XM & \XM & \XM & \XM & \CM & \CM & \CM \\
~ & CVE-2017-6838 & IO & \CM & \XM & \CM & \XM & \XM & \CM & \CM & \CM & \CM \\
\midrule
\multirow{4}*{elfutils} & CVE-2017-7607 & BO & \XM & \CM & \XM & \XM & \CM & \CM & \XM & \XM & \CM \\
~ & CVE-2017-7610 & BO & \XM & \XM & \XM & \XM & \XM & \CM & \CM & \CM & \CM\\
~ & CVE-2017-7611 & BO & \XM & \XM & \XM & \XM & \XM & \CM & \CM & \CM & \CM\\
~ & CVE-2017-7612 & BO & \XM & \XM & \XM & \XM & \XM & \CM & \CM & \CM & \CM\\
\midrule
\multirow{6}*{imageworsener} & CVE-2017-7962 & DZ  & \XM & \CM & \CM & \XM & \XM & \XM & \XM & \XM & \XM\\
~ & CVE-2017-8325 & BO & \XM & \XM & \XM & \XM & \XM & \XM & \XM & \XM & \XM\\
~ & CVE-2017-9204 & BO & \XM & \XM & \XM & \XM
& \XM & \XM & \XM & \XM & \XM\\
~ & CVE-2017-9205 & BO & \XM & \XM & \XM & \XM & \XM & \XM & \XM & \XM & \XM\\
~ & CVE-2017-9206 & BO & \XM & \XM & \XM & \XM & \XM & \XM & \XM & \XM & \XM\\
~ & CVE-2017-9207 & BO & \CM & \XM & \XM & \XM & \XM & \XM & \XM & \XM & \XM\\
\midrule
\multirow{2}*{qpdf} & CVE-2017-9209 & IL & \XM & \XM & \XM & \XM & \XM & \XM & \CM & \CM & \CM \\
~ & CVE-2017-9210 & IL & \XM & \XM & \XM & \XM & \XM & \XM & \XM & \CM & \CM \\
\midrule
\multirow{4}*{ngiflib} & CVE-2018-10677 & BO  & \XM & \XM & \XM & \XM & \XM & \CM & \CM & \XM & \CM \\
~ & CVE-2018-10717 & BO  & \XM & \XM & \XM & \XM & \XM & \CM & \CM & \CM & \CM \\
~ & CVE-2019-16346 & BO & \XM & \XM & \XM & \XM & \XM & \XM & \CM & \XM & \CM \\
~ & CVE-2019-16347 & BO & \XM & \XM & \XM & \XM & \XM & \CM & \CM & \XM & \CM \\
\midrule
\multicolumn{3}{c|}{\textbf{Total}} & 6/55 & 4/55 &
  12/55 & 1/55 & 30/55 & 27/55 & 39/55 & 36/55 &
  \textbf{46/55} \\
\bottomrule
\end{tabular}}
\vspace{-4mm}
\end{table}
\noindent\textbf{Experiment Setup.}
We evaluate \sysname (K.R.), which employs the optimal LLM Gemini-3.1-Pro in RQ1, against eight state-of-the-art AVR approaches described in Section \ref{baselines}: VulRepair \cite{vulrepair} (V.R.), VulMaster \cite{vulmaster} (V.M.), VulnFix \cite{vulnfix} (V.F.), ExtractFix \cite{extractfix} (E.F.), LoopRepair \cite{looprepair} (L.R.), APPatch \cite{appatch} (AP.), PatchAgent \cite{patchagent} (P.A.), and San2Patch \cite{san2patch} (S.A.). Following Hu et al. \cite{DBLP:conf/uss/Hu0SGZXY025}, for two learning-based approaches \cite{vulrepair,vulmaster}, we randomly split the de-duplicated VulRD \cite{vulrepair} dataset, which removes vulnerabilities that are already contained in our benchmark \textsc{Vul4C}, into 80\%-20\% for training and validation, and set the beam size to 50, meaning that 50 candidate patches are produced for each applicable vulnerability. Regarding LLM-based and agentic approaches, we run them using the same backbone model as ours to ensure a fair comparison. In addition, as VulRepair, VulMaster, and APPatch require exact vulnerability manifestation or fix locations, we provide the \emph{diff} version between the vulnerable and patched code as an oracle.

\noindent\textbf{Results.}
Table~\ref{tab:rq2} presents the performance comparison between \sysname and the state-of-the-art baselines on each CVE. A vulnerability is deemed successfully patched (denoted by \CM) only if there exists a patch that passes both the security tests (via PoCs) and functionality tests (via unit test suites); otherwise, it is marked as \XM. Overall, \sysname substantially outperforms all other approaches. In particular, compared to the best-performing baseline PatchAgent, \sysname patches seven more vulnerabilities in total. Meanwhile, \sysname improves over the most related work APPatch, which also follows the RAG paradigm, by 70.37\% (46 vs. 27), indicating the superiority of our proposed knowledge-level retrieval. It is noteworthy that learning-based AVR approaches perform extremely poorly in the majority of cases, even though they have been provided with perfect localization information. For example, the state-of-the-art VulMaster can only patch four out of 55 vulnerabilities, yielding a success rate of just 7.27\%. These results are consistent with previous studies \cite{DBLP:conf/uss/Hu0SGZXY025,DBLP:conf/issta/WuJPLD0BS23,DBLP:journals/corr/abs-2512-22633}, which suggest that learning-based AVR approaches rely heavily on memorized token sequences rather than capturing generalizable repair patterns. In contrast, despite the limited scalability, traditional constraint-based VulnFix can still address 21.82\% (12/55) of the total vulnerabilities.

\begin{figure}[t]
\centering
\begin{subfigure}{0.38\linewidth}
\centering
\includegraphics[width=\linewidth]{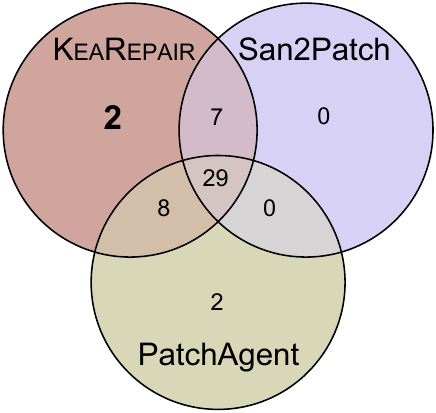}
\caption{vs. agentic approaches.\label{fig:senB}}
\end{subfigure}
\begin{subfigure}{0.55\linewidth}
\centering
\includegraphics[width=\linewidth]{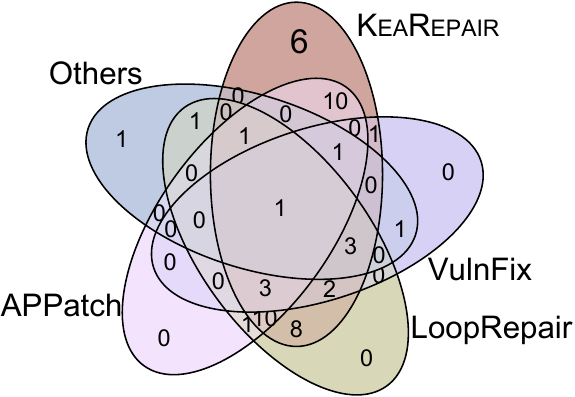}
\caption{vs. other approaches.\label{fig:senA}}
\end{subfigure}
\caption{Overlapping among CVEs fixed by different approaches.\label{fig:venn}}
\vspace{-4mm}
\end{figure}

% \begin{wrapfigure}{r}{.53\linewidth}
%     \vspace{-2mm}
%     % \hspace{-5mm}
%     \centering
%     \includegraphics[width=\linewidth]{Venn.pdf}
%     \caption{Overlapping among CVEs repaired by different approaches.\label{fig:venn}}
%     \vspace{-4mm}
% \end{wrapfigure}
\noindent\textbf{Overlap Analysis.}
To investigate to what extent \sysname complements existing approaches, we further draw a Venn diagram to showcase the performance difference. For a better presentation, we first compare \sysname against two most related agent-based approaches. As shown in Fig. \ref{fig:senB}, while there is substantial overlap in the vulnerabilities resolved by all three approaches, each one also uniquely resolves a small subset of cases except San2Patch. In particular, \sysname can provide the correct patch for nine unique vulnerabilities that the best-performing baseline PatchAgent misses. We also independently illustrate the Top-3 best non-agentic baselines (i.e., LoopRepair, APPatch, and VulnFix) in Fig. \ref{fig:senA} on the basis of the number of correctly repaired vulnerabilities, and divide the rest approaches into one group (named ``Others'') for easy reference. We observe that \sysname is able to fix the most number of unique vulnerabilities of six that prior approaches cannot address. Compare with other two LLM-based approaches, LoopRepair and APPatch, \sysname can repair 18 and 20 unique vulnerabilities, respectively, exemplifying the superiority of \sysname.

\rqbox{\ding{45} \textbf{Answer to RQ2}
    $\blacktriangleright$ The performance improvements of \sysname over the state-of-the-art baselines are significant. Particularly, \sysname outperforms the best-performing PatchAgent by 17.95\% in terms of successfully repaired vulnerabilities, and is able to fix the most number of unique CVEs that prior approaches cannot address. $\blacktriangleleft$}

\begin{table}[t]
\centering
\caption{Ablation study of \sysname components.}
\label{tab:rq4}
\renewcommand\arraystretch{0.8}
\resizebox{\columnwidth}{!} {
\begin{tabular}{lccccccc}
\toprule
\multirow{2.5}{*}{\textbf{Variant}} &
\multicolumn{6}{c}{\textbf{Vulnerability Type}} & \multirow{2.5}{*}{\textbf{Total}}\\
\cmidrule(lr){2-7}
~ & \textbf{BO} & \textbf{IO} & \textbf{DZ} & \textbf{UAF} & \textbf{NPD} & \textbf{IL} & ~\\
\midrule
\sysname (Full) & 30/38 & 6/6 & 4/5 & 2/2 & 2/2 & 2/2 & \textbf{46/55} \\
\midrule
\quad \emph{-w/o Retrieval (CF)}  
& 26/38 & 4/6 & 2/5 & 2/2 & 1/2 & 1/2 & 36/55\\
\quad \emph{-w/o Retrieval (DF)}  
& 24/38 & 4/6 & 3/5 & 1/2 & 2/2 & 1/2 & 35/55\\
\quad \emph{-w/o Retrieval}       
& 22/38 & 3/6 & 2/5 & 2/2 & 1/2 & 0/2 & 30/55 \\
\midrule
\quad \emph{-w/o Diagnosis}       
& 26/38 & 5/6 & 4/5 & 1/2 & 2/2 & 0/2 & 38/55 \\
\quad \emph{-w/o Iteration}       
& 19/38 & 1/6 & 3/5 & 1/2 & 1/2 & 1/2 & 26/55 \\
\bottomrule
\end{tabular}}
\vspace{-4mm}
\end{table}

\subsection{RQ3: Ablation Study}
\noindent\textbf{Experiment Setup.}
To quantify the contribution of each component, we evaluate five ablated variants against the full implementation by disabling one key component at a time.

\begin{itemize}[leftmargin=1em]
    \item \textbf{\emph{w/o Retrieval (CF/DF):}}
    Distills and retrieves the vulnerability knowledge solely from the view of \textbf{D}ata (/\textbf{C}ontrol) \textbf{F}low for patch generation.

    \item \textbf{\emph{w/o Retrieval:}}
    Bypasses the vulnerability knowledge base construction phase, injecting raw diagnostic report directly into the prompt to instruct the repair agent to generate fixes in a zero-shot setting.
        
    \item \textbf{\emph{w/o Diagnosis:}}
    Disables the tool-augmented diagnostic agent, Instructing the LLM to generate root causes from vulnerable code for retrieval and generation.

    \item \textbf{\emph{w/o Iteration:}}
    Replaces the feedback-based iteration mechanism with one-step repair paradigm, outputting the initial patch without further refinement.
\end{itemize}

\noindent\textbf{Results.}
Table~\ref{tab:rq4} summarizes the ablation results across all variants, with the best results highlighted in bold and the largest performance drop in the underline. The experimental results show that the full \sysname configuration achieves the highest performance across all vulnerability types, and removing any single component results in a measurable performance decline. For example, when retrieving demonstrations solely from the view of data flow or control flow, the repair success rate of \sysname drops by 21.74\% (46$\rightarrow$36) and 23.91\% (46$\rightarrow$35), respectively. Once the entire retrieval pipeline is removed, we can observe a substantial performance degradation (46$\rightarrow$30). This confirms that the integration of dual-view vulnerability knowledge is essential for effectively capturing the diversity of vulnerabilities. Notably, we find that the feedback-based iteration mechanism makes the greatest contribution among all variants. Compared to the variant without iteration, \sysname significantly improves the repair rate by 76.92\% (26$\rightarrow$46). The results indicate that single-turn repair attempt is often insufficient for real-world vulnerabilities, and that feedback-driven refinement is critical for producing more plausible patches, which is consistent with recent works \cite{yin2024thinkrepair,looprepair}.

\rqbox{\ding{45} \textbf{Answer to RQ3}
    $\blacktriangleright$ Each component makes a vital and positive contribution to \sysname. The most important component is the feedback-based iteration mechanism that results in 76.92\% improvement in repair success rate. $\blacktriangleleft$}

\subsection{RQ4: Sensitivity Analysis}\label{sens}
\noindent\textbf{Experiment Setup.}
To keep simplicity, $\lambda$ is varied from 0 to 1 with an internal of 0.1. All other configurations are kept consistent with RQ2, including the default LLM backbone and metric computation.

\begin{wrapfigure}{r}{.5\linewidth}
    \vspace{-2mm}
    % \hspace{-5mm}
    \centering
    \includegraphics[width=\linewidth]{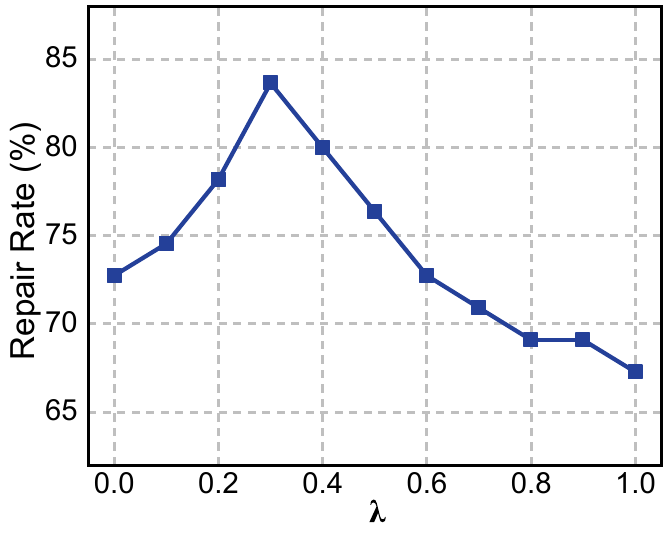}
    \caption{Parameter analysis of the weight coefficient $\lambda$.\label{fig:line}}
    \vspace{-4mm}
\end{wrapfigure}
\noindent\textbf{Results.}
Fig.~\ref{fig:line} presents the performance of \sysname with varying weight coefficient $\lambda$. We can observe that, the repair success rate stably goes up with the increasing of $\lambda$, and reaches the optimal performance at 0.3. Nevertheless, as the proportion of code-level similarity continues to increase, the number of fixes degrades sharply. A possible reason is that project-specific identifiers and incidental logic of lexically-similar examples introduce contextual noise, making LLMs more inclined to mimic the verbatim code patterns of the provided demonstrations, rather than learn high-level repair strategies.

\rqbox{\ding{45} \textbf{Answer to RQ4}
    $\blacktriangleright$ The repair success rate stably goes up with the increasing of $\lambda$, and our default hyper-parameter setting achieve optimal results. $\blacktriangleleft$}

\section{Discussion}\label{discussion}

\subsection{Analysis of Failure Cases}
Despite superior performance, we carefully analyze all the cases where patching fails on the \textsc{Vul4C} dataset, and identify two major failure modes.

\noindent\textbf{Challenges in Static Analysis.}
\sysname relies on static analysis (e.g., pre-computed def-use chains) to collect verified program facts for vulnerability diagnosis. However, real-world C/C++ projects often contain complex macros, compiler-specific extensions, and conditionally compiled paths that are difficult to resolve without full build-system integration. When static analyzers yield imprecise or unsound results, diagnosis falls back from fact-grounded reasoning to under-constrained LLM reasoning. A representative case is CVE-2016-1838 \cite{CVE-2016-1838}, where deeply nested macro expansions prevent reliable scope resolution and intra-procedural data-flow analysis, ultimately resulting in incorrect patches.

\noindent\textbf{Limitations of Single-Location Fixes.}
Another common issue involves inter-procedural vulnerabilities. Despite high-quality, our used \textsc{PrimeVul} dataset constructs vulnerability-patch pairs at the function-level, resulting in distilled vulnerability knowledge fails to capture global context such as cross-file dependencies. As a result, the repair agent may be misled to produce patches that are syntactically valid within the target file but semantically incomplete at the project-level. As revealed in Table \ref{tab:rq2}, \emph{all} six CVEs in \code{imageworsener} require coordinated changes across multiple files, which none of the investigated LLM-based and agentic approaches can handle.

\subsection{Cross-Language Evaluation}
\noindent\textbf{Motivation.}
While demonstrated superior performance, the heavy reliance on benchmarks centered on C/C++ significantly constrains the generalizability of evaluation results. Prior works \cite{DBLP:journals/corr/abs-2508-03470,DBLP:journals/tosem/YangTRZKBGJ26} have shown that LLMs have promising capabilities for multilingual vulnerability repair tasks due to their language-agnostic capabilities. For this purpose, we are interested to investigate whether our proposed \sysname can handle more diverse vulnerabilities beyond C/C++.

\noindent\textbf{Experiment Setup.}
We evaluate the generalizability of \sysname on \textsc{PatchEval} \cite{DBLP:journals/corr/abs-2511-11019}, a multilingual benchmark for Go, JavaScript, and Python. To be specific, we consider a subset of \textbf{173} CVEs equipped with runtime sandbox environments, enabling patch validation through security and/or functionality tests. As specialized vulnerability repair agents, i.e., PatchAgent and San2Patch, only take the sanitizer report as input, we also evaluate another general-purpose approach SWE-Agent \cite{DBLP:conf/nips/YangJWLYNP24}. Other experimental settings, such as LLM backbones and model configurations, are the same as those in RQ1.

\noindent\textbf{Results.}
As shown in Table~\ref{tab:crosslang}, \sysname generalizes effectively to multilingual vulnerability repair tasks. In particular, when paired with Qwen3.5-Plus, \sysname achieves the best performance, patching 148 out of 173 (85.55\%) cases, followed by Claude-Sonnet-4-6 with 143 successful repairs. Additionally, for each LLM, incorporating \sysname yields a significant improvement over its standalone performance as before, with gains ranging from 89.09\% (Gemini-3.1-Pro) to 516.67\% (Qwen3.5-Plus), demonstrating the practical implications of \sysname in real-world scenarios, in which multilingual development is prevalent \cite{DBLP:journals/tosem/LiMYMLC24}. Notably, Gemini-3.1-Pro which outperforms all other models in RQ1 can only patch 104 out of 173 vulnerabilities for a success rate of 60.12\%. These results indicate that there is no ``one-size-fits-all'' model can always perform the best, indicating the necessity of empirically selecting a proper LLM. We also observe that integrating general-purpose SWE-Agent does not lead to performance gains, even degrades results, underscoring the importance of developing domain-specialized agents tailored.

\begin{table}[t]
\centering
\caption{Performance on the \textsc{PatchEval} dataset.}
\label{tab:crosslang}
% \setlength{\tabcolsep}{4pt}
% \small
\renewcommand\arraystretch{0.8}
\setlength{\tabcolsep}{0.5\tabcolsep}
\resizebox{\columnwidth}{!} {
\begin{tabular}{clrrrr|rrrr}
\toprule
\multirow{2.5}{*}{\textbf{Arch.}} & \multirow{2.5}{*}{\textbf{Backbone Model}} &
\multicolumn{4}{c|}{\textbf{PoC-only}} &
\multicolumn{4}{c}{\textbf{PoC \& Unit}} \\
\cmidrule(lr){3-6} \cmidrule(lr){7-10}
~  & ~ & \textbf{Go} & \textbf{Js} & \textbf{Py} & \textbf{Total}
& \textbf{Go} & \textbf{Js} & \textbf{Py} & \textbf{Total} \\
\midrule
\multirow{5}*{\rotatebox{90}{Standalone}} & GPT-5.1            & 11/62 & 13/47 &  7/64 & 31/173
                       & 10/62 & 10/47 &  6/64 &  26/173  \\
~ & Claude-Sonnet-4-6  & 20/62 & 19/47 & 10/64 & 49/173
                       & 18/62 & 13/47 &  9/64 &  40/173  \\
~ & Gemini-3.1-Pro     & 19/62 & 23/47 & 17/64 & \textbf{59/173}
                       & 19/62 & 20/47 & 16/64 &  \textbf{55/173}  \\
~ & Qwen3.5-Plus       & 10/62 & 13/47 &  8/64 & 31/173
                       & 10/62 & 9/47 &  5/64 &  24/173  \\
~ & DeepSeek-V3.2      & 14/62 & 14/47 & 10/64 & 38/173
                       & 14/62 & 10/47 &  8/64 &  32/173  \\
\midrule
\multirow{5}*{\rotatebox{90}{\sysname}} &  GPT-5.1            & 49/62 & 43/47 & 52/64 & 144/173 & 43/62 & 40/47 & 47/64 & 130/173\\
~ & Claude-Sonnet-4-6  & 57/62 & 44/47 & 54/64 & 155/173
                       & 56/62 & 38/47 & 49/64 & 143/173 \\
~ & Gemini-3.1-Pro     & 38/62 & 39/47 & 53/64 & 130/173
                       & 36/62 & 30/47 & 38/64 & 104/173\\
~ & Qwen3.5-Plus       & 58/62 & 45/47 & 52/64 & \textbf{155/173}
                       & 56/62 & 43/47 & 49/64 & \textbf{148/173}\\
~ & DeepSeek-V3.2      & 33/62 & 34/47 & 43/64 & 110/173
                       & 29/62 & 30/47 & 40/64 &  99/173 \\
\midrule
\multirow{5}*{\rotatebox{90}{SWE-Agent}} &  GPT-5.1            & 12/62 & 17/47 & 17/64 & 46/173
                       & 10/62 & 15/47 & 14/64 &  39/173 \\
~ & Claude-Sonnet-4-6  & 14/62 & 17/47 & 12/64 & 43/173
                       & 14/62 & 16/47 & 11/64 &  41/173 \\
~ & Gemini-3.1-Pro     & 14/62 & 21/47 & 22/64 & \textbf{57/173}
                       & 13/62 & 20/47 & 21/64 &  \textbf{54/173} \\
~ & Qwen3.5-Plus       & 11/62 & 19/47 & 15/64 & 45/173
                       &  9/62 & 15/47 & 13/64 &  37/173 \\
~ & DeepSeek-V3.2      & 13/62 & 19/47 & 16/64 & 48/173
                       & 13/62 & 17/47 & 15/64 &  45/173 \\
\bottomrule
\end{tabular}
}
\vspace{-4mm}
\end{table}

\subsection{Threats to Validity}
\noindent\textbf{Threats to External Validity} come from the risk of data leakage. We evaluated the performance of \sysname on 55 and 173 reproducible vulnerabilities sourced from \textsc{Vul4C} and \textsc{PatchEval}, respectively. Since LLMs are trained on massive inscrutable corpora, it is difficult to rule out the possibility that CVEs or their patches in these static datasets are included in their pre-training data, resulting in artificially inflated performance. To mitigate this threat, we plan to construct a contamination-resistant, reproducible, and continuously updatable benchmark tailored to real-world AVR tasks. Despite that, our controlled studies under the same LLM backbone demonstrate that the observed improvements are attributable to \sysname's design rather than differences in training data.

\noindent\textbf{Threats to Internal Validity} refer to the generalizability of our findings. We evaluated \sysname on four popular programming languages, i.e., C/C++, Python, JavaScript, and Go, and thus our experimental results may not generalizable to other programming languages. However, we believe the key unique design of \sysname (i.e., performing tool-augmented agentic reasoning to collect verified program facts for reliable root cause analysis, and improving the patch quality via dual-view knowledge retrieval) is language-agnostic, and thus can be easily ported for a new programming language.

\noindent\textbf{Threats to Construct Validity} arise from the sufficiency of our oracle-based evaluation measures. Following prior work \cite{DBLP:journals/corr/abs-2511-11019}, we combine security tests and functionality tests to confirm whether a vulnerability is fixed. Unfortunately, these dynamic checks still cannot guarantee the correctness of generated patches as they may pass all available tests while still introducing regressions not covered by the test suites \cite{DBLP:conf/issta/WeiCZSM24}. Manual validation is preferred, yet it is knowledge-intensive. Thus, automated approaches capable of assessing what a patch fixes, not whether the observed crash is suppressed are required.

\section{Conclusion}\label{Conclusion}
Frontier foundation models are discovering vulnerabilities at a pace and depth that human red teams simply cannot match, resulting in a faster-growing backlog of unresolved critical issues. In this paper, we presented \sysname, a knowledge-enhanced agentic approach for automated vulnerability repair. By combining vulnerability knowledge base construction, fact-grounded vulnerability diagnosis, and retrieval-augmented patch generation, \sysname grounds LLM-based repair in verified program facts and reusable historical fix knowledge. 
% This design mitigates the limitations of purely static LLM reasoning and improves the relevance of retrieved exemplars for patch synthesis. 
Extensive experiments on real-world vulnerabilities demonstrate that \sysname substantially outperforms state-of-the-art baselines in both repair effectiveness and cross-language generalizability.

% \clearpage
\balance

\bibliographystyle{./bibliography/IEEEtran}
\bibliography{./bibliography/IEEEabrv,./bibliography/IEEEexample}

\end{document}